\RequirePackage{fixltx2e}
\documentclass[twocolumn,floatfix,showpacs,preprintnumbers,
nofootinbib,superscriptaddress]{revtex4-1}
\usepackage[english]{babel}
\usepackage[utf8x]{inputenc}
\usepackage{mathrsfs}
\usepackage{amssymb}
\usepackage{amsmath}
\usepackage{mathtools}
\usepackage{mathrsfs}
\usepackage{graphicx}
\usepackage{fixltx2e}

\usepackage{xcolor}
\definecolor{lcolor}{rgb}{0.5,0,0}
\definecolor{citcolor}{rgb}{0,0.3,0.0}

\usepackage[breaklinks,colorlinks,urlcolor=blue,citecolor=citcolor,linkcolor=lcolor]{hyperref}
\usepackage{subdepth}

\newcommand{\rt}{{\mathbf{r}}}
\newcommand{\xt}{{\mathbf{x}}}

\newcommand{\ktt}{k_\perp} 

\newcommand{\ud}{\mathrm{d}}

\newcommand{\tr}{\, \mathrm{Tr} \, }

\newcommand{\nc}{{N_\mathrm{c}}}
\newcommand{\nf}{{n_\mathrm{f}}}

\newcommand{\cf}{C_\mathrm{F}}

\newcommand{\nr}[1]{(\ref{#1})}

\newcommand{\qso}{Q_\mathrm{s,0}}

\newcommand{\lqcd}{\Lambda_{\mathrm{QCD}}}
\newcommand{\as}{\alpha_{\mathrm{s}}}

\newcommand{\eq}{Eq.~}

\newcommand{\eqs}{Eqs.~}

\newcommand{\xbj}{{x_{Bj}}}

\newcommand{\kcal}{\mathcal{K}}

\def\figscale{0.43}
\def\figspace{0mm}

\begin{document}

\author{B. Duclou\'e}
\affiliation{
Department of Physics, 
 P.O. Box 35, 40014 University of Jyv\"askyl\"a, Finland
}
\affiliation{
Helsinki Institute of Physics, P.O. Box 64, 00014 University of Helsinki,
Finland
}
\author{H. Hänninen}
\affiliation{
Department of Physics, 
 P.O. Box 35, 40014 University of Jyv\"askyl\"a, Finland
}
\author{T. Lappi}
\affiliation{
Department of Physics, 
 P.O. Box 35, 40014 University of Jyv\"askyl\"a, Finland
}
\affiliation{
Helsinki Institute of Physics, P.O. Box 64, 00014 University of Helsinki,
Finland
}
\author{Y. Zhu}
\affiliation{
Department of Physics, 
 P.O. Box 35, 40014 University of Jyv\"askyl\"a, Finland
}
\affiliation{
Helsinki Institute of Physics, P.O. Box 64, 00014 University of Helsinki,
Finland
}

\title{Deep inelastic scattering in the dipole picture at next-to-leading order}


\preprint{}

\begin{abstract}
We study quantitatively the importance of the recently derived NLO corrections to the DIS structure functions at small~$x$ in the dipole formalism. We show that these corrections can be significant and depend on the factorization scheme used to resum large logarithms of energy into renormalization group evolution with the BK equation.  This feature is similar to what has recently been observed for single inclusive forward hadron production. Using a factorization scheme consistent with the one recently proposed for the single inclusive cross section, we show that it is possible to obtain meaningful results for the DIS cross sections.
\end{abstract}

\maketitle

\section{Introduction}

At high energy (or equivalently small values of the longitudinal momentum fraction $x$), the gluon density in hadrons can become nonperturbatively large: this is the regime of gluon saturation. However, the evolution of this gluon density as a function of the momentum fraction $x$ can still be computed using weak coupling techniques, leading to the Balitsky-Kovchegov (BK) evolution equation~\cite{Balitsky:1995ub,Kovchegov:1999yj}. Knowing the initial gluon density at a given $x=x_0$, one can thus evolve it perturbatively to any $x<x_0$. This initial condition involves nonperturbative dynamics and needs to be extracted from data, but the evolution equation then gives a first principles prediction for smaller $x$.

The cleanest process to study the partonic structure of hadrons is provided by deep inelastic scattering (DIS). At small~$x$ this process is most conveniently understood in the dipole picture, where the scattering is factorized into a QED splitting of the virtual photon into a quark-antiquark dipole, and the subsequent QCD interaction of this dipole with the target. Here the BK equation describes the dependence of the dipole-target scattering amplitude on the collision energy. Several groups have been able to obtain satisfactory fits to HERA 
DIS data in the leading order dipole picture, using the BK equation with running coupling corrections (see for example Refs.~\cite{Albacete:2010sy,Lappi:2013zma}). To advance the saturation formalism to next-to-leading order, two key ingredients are needed: the NLO BK equation and the process-dependent NLO impact factors. In addition to many recent methodological developments for these higher order calculations (see e.g. \cite{Lappi:2016oup,Ayala:2017rmh}), progress has been made in both of these directions. The NLO corrections to the BK equation have been computed in~\cite{Balitsky:2008zza} and evaluated numerically in~\cite{Lappi:2015fma}, where it was shown that they can lead to unphysical results. This problem has been subsequently solved by resumming classes of large logarithms~\cite{Beuf:2014uia,Iancu:2015vea,Iancu:2015joa}, indeed leading to reasonable results~\cite{Lappi:2016fmu}.

Concerning impact factors, most of the recent work has concentrated on the NLO corrections to single inclusive forward hadron production. The impact factor for this process has been known for some time~\cite{Chirilli:2011km,Chirilli:2012jd}, but the first numerical implementation of these expressions showed that they can make the cross section negative when the transverse momentum of the produced hadron is of the order of a few GeV~\cite{Stasto:2013cha}. Several works have been devoted to solving this issue~\cite{Kang:2014lha,Stasto:2014sea,Altinoluk:2014eka,Watanabe:2015tja,Ducloue:2016shw}, and recently a new proposed formulation of the NLO cross section~\cite{Iancu:2016vyg} was shown to lead to physical results~\cite{Ducloue:2017mpb}, albeit with a remaining issue concerning the best way to implement a running QCD coupling constant. 

Also the impact factor for DIS in the dipole picture has been studied in several papers~\cite{Balitsky:2010ze,Beuf:2011xd,Boussarie:2014lxa,Boussarie:2016ogo}. However, the full expressions in the mixed space representation (longitudinal momentum, but transverse coordinate) that are most naturally combined with BK evolution have only become available more recently~\cite{Beuf:2016wdz,Beuf:2017bpd}. For a practical implementation of these results it is essential to match the impact factor calculation with the evolution equation in the correct way, i.e. to factorize the leading high energy logarithms into the high energy evolution. As we shall discuss below, the situation here is very analogous to that of single inclusive particle production.

The main purpose of this paper is twofold. We firstly want to study the importance of the NLO  corrections to have a first estimate of the stability of the perturbative expansion for this quantity. Secondly we want to develop a good factorization procedure for matching the renormalization group evolution with the previous calculation of the impact factor. Both of these are prerequisites  for a description of experimental data, which will be pursued in a continuation of this work.  Our focus in this paper is to demonstrate the feasibility of the factorization scheme and study the general characteristics of the NLO corrections to the cross sections. A full NLO calculation will additionally require including an NLO evolution equation. In this paper we shall first, in Sec.~\ref{sec:eqs}, briefly present the NLO impact factor as calculated in~\cite{Beuf:2016wdz,Beuf:2017bpd}. We shall then, in Sec.~\ref{sec:results} quantify the effects of the NLO corrections for the $Q^2$ and $\xbj$-dependence of the transverse and longitudinal DIS cross sections.

\section{Impact factor}
\label{sec:eqs}

In the dipole framework, the interaction of a virtual photon with  the proton in DIS is factorized as the scattering of a quark-antiquark dipole with the proton. 
At leading order, the expressions for the cross sections of transversally or longitudinally polarized virtual photons $\sigma_{L,T}$ read
\begin{align}\label{eq:lo}
\sigma_{L,T}^\text{LO}(\xbj,Q^2) & = 4 \nc \alpha_{em} \sum_f e_f^2 \int_0^1 \ud z_1 \nonumber \\
& \quad \times \int_{\xt_0, \xt_1} \kcal_{L,T}^\text{LO}(z_1,\xt_0,\xt_1,\xbj),
\end{align}
with the shorthand $\int_{\xt_0} =\int\frac{\ud^2 \xt_0}{2\pi}$. The integrands are given by the squares of the light cone wave functions for the $\gamma^*\to q\bar{q}$ splitting  and the scattering amplitudes for the $q\bar{q}$ dipole to scatter off the target
\begin{align}
\kcal_{L}^\text{LO}(z_1,\xt_0,\xt_1,X) &= 4 Q^2 z_1^2 (1-z_1)^2 \nonumber \\
& \quad \times K_0^2(Q X_2) \left(1-S_{01}(X)\right), \\
\kcal_{T}^\text{LO}(z_1,\xt_0,\xt_1,X) &= Q^2 z_1 (1-z_1) \left(z_1^2+(1-z_1)^2\right) \nonumber \\
& \quad \times K_1^2(Q X_2) \left(1-S_{01}(X)\right),
\end{align}
for the longitudinal ($L$) and transverse ($T$) polarized virtual photon respectively. Here the argument of the Bessel functions, related to the lifetime of the  $q\bar{q}$ fluctuation, is
$
X_2^2=z_1(1-z_1)\xt_{01}^2 .
$
The scattering amplitude of the dipole is given, in the CGC picture, by the two point function of a correlator of Wilson lines, namely
\begin{align}
S_{01}(X) & \equiv S(\xt_{01}=\xt_0-\xt_1,X) \nonumber \\
&=\left< \frac{1}{\nc}\tr U(\xt_0)U^\dag(\xt_1) \right>_X,
\end{align}
where we denote by $X$ the momentum fraction (corresponding to the evolution variable in the BK equation $y=\ln 1/X$) at which the Wilson line correlator is to be evaluated.

The NLO corrections to these expressions have been computed in~\cite{Beuf:2016wdz,Beuf:2017bpd}. They involve two kinds of terms: the one loop corrections to the $q\bar{q}$-state and a new $q\bar{q}g$-component in the $\gamma^*$ Fock state.  Following the general idea exposed in Ref.~\cite{Iancu:2016vyg} for single inclusive hadron production, we write the (unsubtracted) NLO cross sections as
\begin{equation}
\label{eq:NLO_bare}
\sigma_{L,T}^\text{NLO}
=\sigma_{L,T}^{(0)}
+\sigma_{L,T}^{qg}
+\sigma_{L,T}^\text{dip} \, .
\end{equation}
In this expression, the first term corresponds to the lowest order contribution with an unevolved target (i.e. evaluated at the rapidity $X=x_0$). The terms proportional to $\as$  have been organized into two parts. Firstly the gluon contribution $\sigma_{L,T}^{qg}$ includes all the real contributions (with a gluon emitted into the final state) and a subset of the virtual corrections that need to be combined with the real corrections to cancel any ultraviolet or collinear divergences. The dipole contribution $\sigma_{L,T}^\text{dip}$ contains the rest of the virtual corrections. The separation between these two terms is not unique, but the sum of the two is fully determined by the NLO calculation. The expressions for these terms can be written as
\begin{align}
&\sigma_{L,T}^{qg}
= 8 \nc \alpha_{em} \frac{\alpha_s \cf}{\pi} \sum_f e_f^2 \int_0^1 \ud z_1 \int^{1-z_1} \frac{\ud z_2}{z_2} \nonumber \\
& \hspace{1.1cm} \times \! \int_{\xt_0, \xt_1, \xt_2} \!\! \kcal_{L,T}^\text{NLO}\left(z_1,z_2,\xt_0,\xt_1,\xt_2,X(z_2)\right) , \\
&\sigma_{L,T}^\text{dip}
= 4 \nc \alpha_{em} \frac{\alpha_s \cf}{\pi} \sum_f e_f^2 \int_0^1 \ud z_1 \nonumber \\
& \times \! \int_{\xt_0, \xt_1} \!\! \kcal_{L,T}^\text{LO}(z_1,\xt_0,\xt_1,X^\text{dip}) \! \left[\frac{1}{2}\ln^2\!\left(\!\frac{z_1}{1\!-\!z_1}\!\right)\!-\!\frac{\pi^2}{6}\!+\!\frac{5}{2}\right],
\label{eq:NLO_dip}
\end{align}
with
\begin{widetext}
\begin{align}
&\kcal_L^\text{NLO}(z_1,z_2,\xt_0,\xt_1,\xt_2,X)= 4 Q^2 z_1^2 (1-z_1)^2 \bigg\{ \! P \! \left(\frac{z_2}{1-z_1}\right) \! \frac{\xt_{20}}{\xt_{20}^2} \!\cdot\! \left(\frac{\xt_{20}}{\xt_{20}^2}-\frac{\xt_{21}}{\xt_{21}^2}\right) \! \left[ K_0^2(Q X_3) \left(1-S_{012}(X)\right)-(\xt_2 \to \xt_0) \right] \nonumber \\
& \hspace{7cm}+ \left(\frac{z_2}{1-z_1}\right)^2 \frac{\xt_{20} \cdot \xt_{21}}{\xt_{20}^2 \xt_{21}^2} K_0^2(Q X_3) \left(1-S_{012}(X)\right) \bigg\} , \\
&\kcal_T^\text{NLO}(z_1,z_2,\xt_0,\xt_1,\xt_2,X)= Q^2 z_1(1-z_1) \nonumber \\
&\hspace{1cm}\times \bigg\{ 
P\left(\frac{z_2}{1-z_1}\right)\left(z_1^2+(1-z_1)^2\right) \frac{\xt_{20}}{\xt_{20}^2} \cdot \left(\frac{\xt_{20}}{\xt_{20}^2}-\frac{\xt_{21}}{\xt_{21}^2}\right) \Big[K_1^2(Q X_3) \left(1-S_{012}(X)\right)-(\xt_2 \to \xt_0) \Big] \nonumber \\
&\hspace{1.6cm} + \left(\frac{z_2}{1-z_1}\right)^2 \left[ \left(z_1^2+(1-z_1)^2\right) \frac{\xt_{20} \cdot \xt_{21}}{\xt_{20}^2 \xt_{21}^2} + 2 z_0 z_1 \frac{\xt_{20} \cdot \xt_{21}}{\xt_{20}^2 X_3^2} - \frac{z_0(z_1+z_2)}{X_3^2} \right] K_1^2(Q X_3) \left(1-S_{012}(X)\right) \bigg\} .
\end{align}
\end{widetext}
Here the longitudinal momentum fractions of the quark, antiquark and gluon are denoted as $z_0,z_1,z_2$ with $z_0+z_1+z_2=1$. The argument of the Bessel functions, related to the lifetime of the $q\bar{q}g$-fluctuation, is $X_3^2 =z_0 z_1 \xt_{01}^2+z_0 z_2\xt_{20}^2+z_2 z_1\xt_{21}^2 $,  $P(z) =1+(1-z)^2$ and the Wilson line operator corresponding to the scattering of the $q\bar{q}g$ state is
\begin{align}
S_{012}(X) & =\frac{\nc}{2\cf}\left(S_{02}(X)S_{21}(X)-\frac{1}{\nc^2}S_{01}(X)\right).
\end{align}

It is important to note that because the functions $\kcal_{L,T}^\text{NLO}\left(z_1,z_2,\xt_0,\xt_1,\xt_2,X\right)$ approach a non-zero value when $z_2 \to 0$ at fixed $X$, the integral over $z_2$ in $\sigma_{L,T}^{qg}$ produces a large logarithm which should be resummed in the BK evolution of the target. We will do this using the same procedure introduced in~\cite{Beuf:2014uia,Iancu:2016vyg} and demonstrated in~\cite{Ducloue:2017mpb} for the case of single inclusive particle production in forward proton-nucleus collisions. Note that, similar to the ``$\cf$-term'' in the case of the single inclusive cross section, the ``dipole''-term does not generate such a large logarithmic contribution and therefore does not contribute to the BK evolution.

The starting point of the BK-factorization procedure is to  identify the first term in Eq.~(\ref{eq:NLO_bare}) as the initial condition for the BK evolution with the longitudinal momentum fraction $x_0\sim 0.01$, i.e.
\begin{equation}
\sigma_{L,T}^\text{IC}
=4 \nc \alpha_{em} \sum_f e_f^2
 \int_0^1 \ud z_1 \int_{\xt_0, \xt_1}\kcal_{L,T}^\text{LO}(z_1,\xt_0,\xt_1,x_0).
\end{equation}
As discussed in great detail in~\cite{Beuf:2014uia,Iancu:2016vyg}, the essential feature required for a stable perturbative expansion is that the dipole correlators in $\sigma_{L,T}^{qg}$ must be evaluated at a rapidity scale that depends on the longitudinal momentum of the emitted gluon, i.e. $z_2$. Here, there are several different possibilities, which  are all equivalent at the leading logarithmic level. At NLO accuracy the different schemes lead to different expressions which are in principle equivalent, but more naturally lend themselves to different approximations. 

The choice advocated in Ref.~\cite{Beuf:2014uia} is to consistently use the probe longitudinal momentum $k^+$ as the evolution variable, sometimes referred to as ``probe evolution''. In this case the evolution rapidity is by definition $y = \ln 1/z_2 + y_0$ with some constant $y_0$ used to make $y=0$ correspond to the initial condition for the evolution. To determine the lower integration limit for $z_2$ in this scheme we have to compare the longitudinal momentum of the emitted soft gluon $z_2q^+$ to momentum scales in the target. The typical target hadronic momentum scale is given by $P^+ = Q_0^2/(2P^-)$, where $Q_0$ is some hadronic low transverse momentum scale and the total target light cone energy $P^-$ is obtained from the total center of mass energy of the $\gamma^*$-target system by $W^2= 2 q^+P^-$.  For the eikonal approximation to be valid we require that the probe gluon momentum is larger than the target momentum scale by a large factor $1/x_0$,  i.e. $z_2 q^+ > (1/x_0) P^+$. This translates, using $\xbj \approx Q^2/W^2$, into an integration limit $z_2 > (\xbj/x_0) (Q_0^2/Q^2)$. If now the soft gluon has a transverse momentum $\ktt$, the light cone energy required from the target to put the $q\bar{q}g$-state on shell is $\Delta k^- \gtrsim \ktt^2/(2z_2 q^+)$. The limit on $z_2$ means that we allow the $\gamma^*$ system to take a fraction $\Delta k^-/P^- \lesssim x_0 (\ktt^2/Q_0^2)$ of the target light cone energy. If the typical gluon $\ktt$ is at the hadronic scale $Q_0$, this is indeed the limit $\Delta k^-/P^- < x_0$ that we would want for the fraction of the target light cone energy. However, the contribution from  $\ktt^2 \sim Q^2 \gg Q_0^2$ goes to larger values of the target momentum fraction $\Delta k^-/P^-$ than we would want. This can generally be expected to be a problem that must be corrected by imposing an additional ``kinematical constraint'' on the evolution equation~\cite{Motyka:2009gi,Beuf:2014uia} and on the impact factor~\cite{Stasto:2014sea,Watanabe:2015tja,Ducloue:2016shw}.

The other option to probe evolution  is to take the view that the evolution variable should always be the target momentum fraction, i.e. the fraction of the target light cone energy $X = \Delta k^-/P^-$. Keeping this momentum fraction small, $X<x_0$, removes the need for an additional kinematical constraint, significantly simplifying the evolution equation. On the other hand using $\Delta k^-/P^-$ as the evolution variable adds the significant complication that this momentum fraction depends on the transverse momentum of the gluon, $X(z_2) \approx \ktt^2/(z_2 W^2)$, and when $z_2$ is not very small also on the momenta of the quark and antiquark. This makes it difficult to implement a light cone energy factorization scale or evolution variable exactly. Parametrically, the transverse momentum $\ktt$ can range from a hadronic scale $Q_0$ to the hard scale $Q$. If one estimates the typical target momentum fraction $\Delta k^-$ assuming
that the typical gluon transverse momentum  is at the  hadronic scale $\ktt^2\sim Q_0^2$, one recovers the same limit $z_2 > (\xbj/x_0) (Q_0^2/Q^2)$ as argued from using $k^+$ as the factorization variable. In contrast, the argument used in the recent work on single inclusive particle production in proton-nucleus collisions~\cite{Iancu:2016vyg,Ducloue:2017mpb} was that, at least in that case, the typical transverse momentum of the gluon in the impact factor is in fact the hard scale of the process $\ktt \sim Q$. Assuming that this is the case also for DIS means that one should restrict the integrals to a smaller phase space $z_2 > (\xbj/x_0)$. This latter is the limit that we will use in this work. In terms of the $k^+$-momentum this limit corresponds to the emitted gluon having longitudinal momentum $z_2 q^+ \gtrsim (Q^2/Q_0^2)(1/x_0) P^+$ instead of the $z_2 q^+ > (1/x_0) P^+$ that one would use in the factorization scheme with $k^+$. This approximation leads to a rather simple formulation for the cross section. Improving the accuracy would require including the additional phase space  $  (\xbj/x_0) (Q_0^2/Q^2)< z_2 < (\xbj/x_0) $ in the cross section on one hand, but cutting out the large logarithmic increase from this region by using a kinematical constraint in the evolution equation, as advocated e.g. in \cite{Beuf:2014uia,Beuf:2016wdz,Beuf:2017bpd}.  Due to the considerably increased complication of this formulation, we will defer studying this alternative to future work.

\begin{figure*}[tbp]
\includegraphics[scale=\figscale]{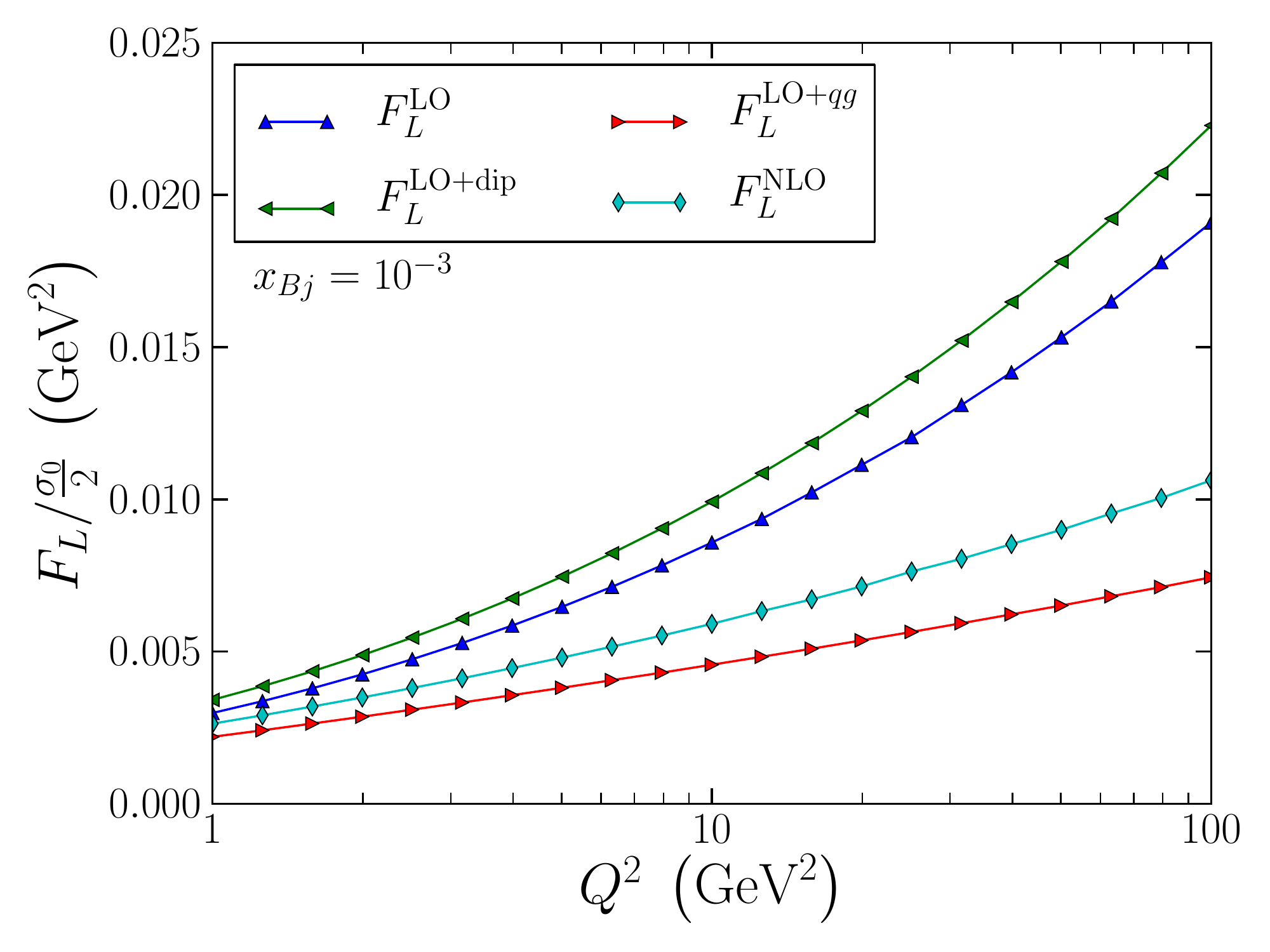}
\hspace{\figspace}
\includegraphics[scale=\figscale]{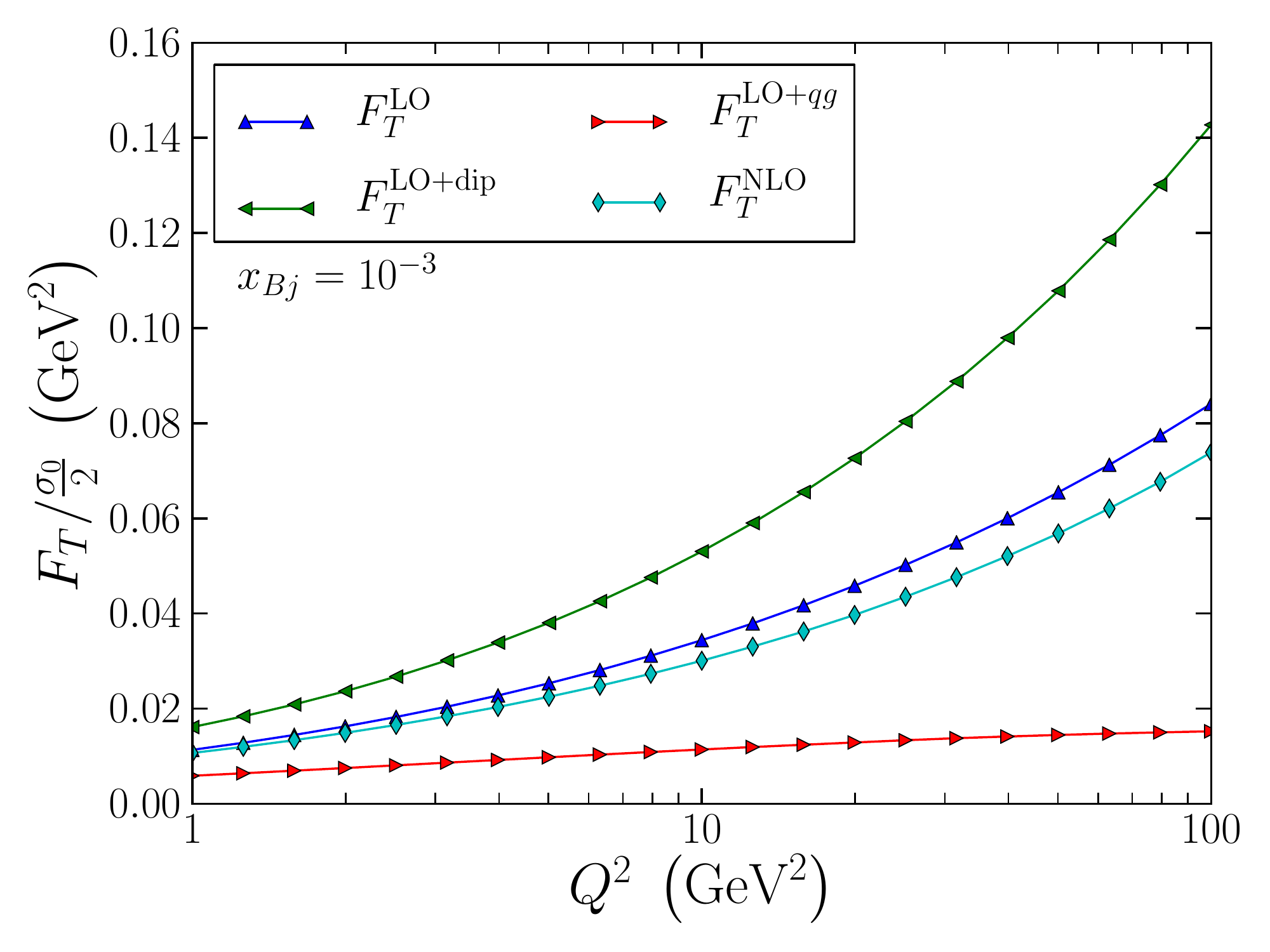}
\caption{LO and NLO contributions to $F_L$ (left) and $F_T$ (right) as a function of $Q^2$ at $\xbj=10^{-3}$ with $\as=0.2$. 
}
\label{fig:fc_Q}
\end{figure*}

\begin{figure*}[tbp]
\includegraphics[scale=\figscale]{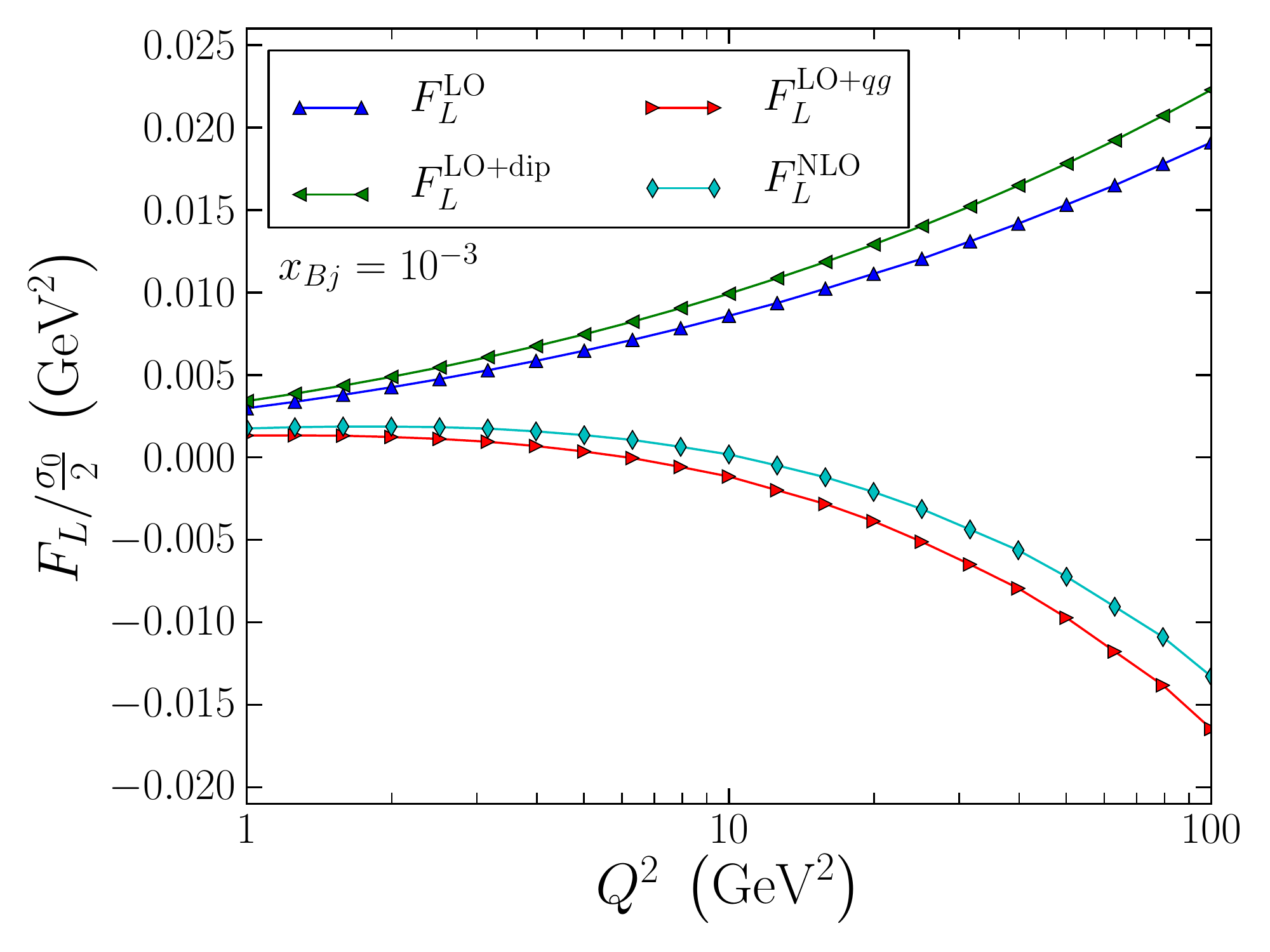}
\hspace{\figspace}
\includegraphics[scale=\figscale]{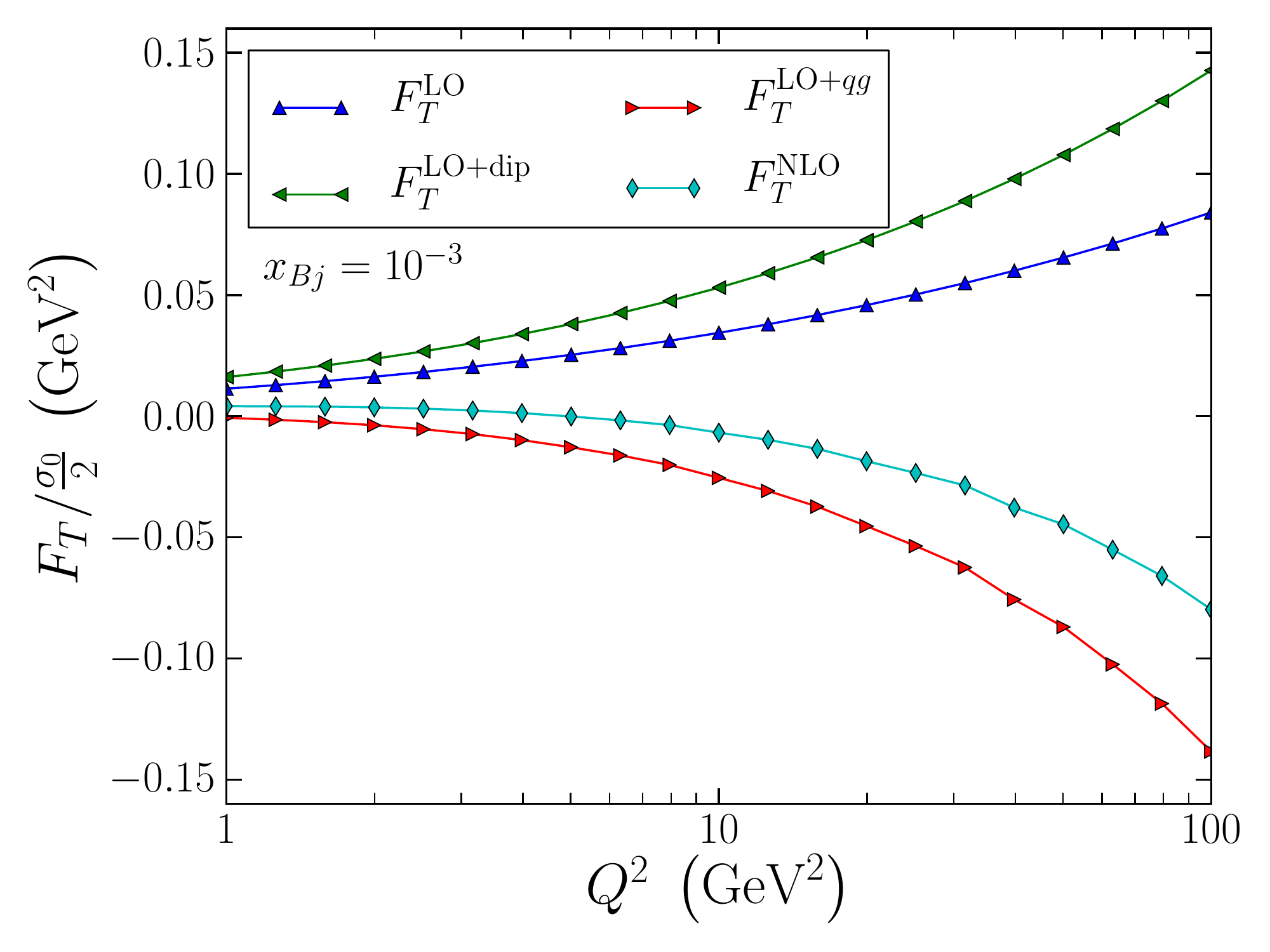}
\caption{LO and NLO contributions to $F_L$ (left) and $F_T$ (right) as a function of $Q^2$ at $\xbj=10^{-3}$ with $\as=0.2$ and using the $\xbj$-subtraction procedure. }
\label{fig:fc_Q_sub}
\end{figure*}

To summarize, in this paper we will follow the choice made for single inclusive particle production in proton-nucleus in~\cite{Iancu:2016vyg,Ducloue:2017mpb}, and choose the target momentum fraction as the evolution variable, supplemented with the assumption that all transverse momenta are of the order $Q$. Thus we take $X(z_2)=\xbj/z_2$ and set the kinematical limit by requiring $X(z_2)< x_0$, i.e. $z_2>\xbj/x_0$. Implementing this limit we can now complete the ``unsubtracted'' form of the cross section (\ref{eq:NLO_bare}) with the lower integration limit in $z_2$ as
\begin{equation}
\label{eq:NLO_unsub}
\sigma_{L,T}^\text{NLO}
=\sigma_{L,T}^\text{IC}
+\sigma_{L,T}^{qg, \text{unsub.}}
+\sigma_{L,T}^\text{dip} \, ,
\end{equation}
with
\begin{align}
\label{eq:NLO_qg_unsub}
\sigma_{L,T}^{qg, \text{unsub.}} &= 8 \nc \alpha_{em} \frac{\alpha_s \cf}{\pi} \sum_f e_f^2 \int_0^1 \ud z_1 \int_{\xbj/x_0}^{1-z_1} 
\frac{\ud z_2}{z_2} \nonumber \\
& \quad \times \int_{\xt_0, \xt_1, \xt_2} \kcal_{L,T}^\text{NLO}\left(z_1,z_2,\xt_0,\xt_1,\xt_2,X(z_2)\right) .
\end{align}
We then note  that taking $z_2=0$ as the explicit $z_2$-argument in $\kcal_{L,T}^\text{NLO}$ (but not in the implicit dependence through $X(z_2)$)  leads to  an integral version of the BK equation. Using this we can also rewrite Eq.~(\ref{eq:NLO_unsub}) in a  form that involves the leading order cross sections with BK-evolved dipole operators evaluated at the scale $\xbj$ instead of $x_0$. The result is a strictly equivalent ``subtracted'' form of the cross section
\begin{equation}
\label{eq:NLO_sub}
\sigma_{L,T}^\text{NLO}
=\sigma_{L,T}^\text{LO}
+\sigma_{L,T}^{qg, \text{sub.}}
+\sigma_{L,T}^\text{dip} \, ,
\end{equation}
where $\sigma_{L,T}^\text{LO}$ is the well known leading order expression \nr{eq:lo} and
\begin{align}
\label{eq:NLO_qg_sub}
& \sigma_{L,T}^{qg, \text{sub.}}
=  8 \nc \alpha_{em} \frac{\alpha_s \cf}{\pi} \sum_f e_f^2  \int_0^1 \ud z_1 \int_{\xbj/x_0}^{1} \frac{\ud z_2}{z_2} \nonumber \\
& \times \! \int_{\xt_0, \xt_1, \xt_2} \! \bigg[ \theta(1\!-\!z_1\!-\!z_2)
 \kcal_{L,T}^\text{NLO}\left(z_1,z_2,\xt_0,\xt_1,\xt_2,X(z_2)\right) \nonumber \\ 
& \hspace{1.9cm} -\kcal_{L,T}^\text{NLO}\left(z_1,0,\xt_0,\xt_1,\xt_2,X(z_2)\right)\bigg] .
\end{align}

Contrary to $\sigma_{L,T}^{qg}$, the dipole term $\sigma_{L,T}^\text{dip}$ is not associated with the rapidity evolution of the target, thus the rapidity scale of the dipole operators in this term is left unspecified. As presented in \cite{Beuf:2016wdz,Beuf:2017bpd}, this term is already integrated over $z_2$. Therefore it is not possible to evaluate the dipole operators in this term at the same scale $X(z_2)=\xbj/z_2$ as in $\sigma_{L,T}^{qg}$, which would arguably be the most natural thing to do. Here we will evaluate this term at $X^\text{dip}=\xbj$ since the integrand vanishes when $z_2 \to 0$ and therefore one can expect the integral to be dominated by the region where $z_2$ is close to 1. Note, however, that the difference between $X=\xbj/z_2$ and $X=\xbj$, while formally subleading for the ``dipole'' term, could be numerically important, as is the case for the analogous $\cf$-terms in single inclusive particle production~\cite{Ducloue:2017mpb}.

To obtain the previous expressions, we followed closely the original idea of Ref.~\cite{Iancu:2016vyg}, which was shown in~\cite{Ducloue:2017mpb} to lead to reasonable numerical results for single inclusive particle production at all transverse momenta. Bear in mind that the two expressions in \eqs\nr{eq:NLO_unsub} and \nr{eq:NLO_sub} are completely equivalent, and are related through the BK evolution equation.
In the following, it will also be interesting to compare the results obtained in this formulation with what we denote here as the ``$\xbj$-subtraction'' scheme, which is expressed as
\begin{equation}
\label{eq:NLO_xbjsub}
\sigma_{L,T}^{\text{NLO},\xbj-\text{sub.}}
=\sigma_{L,T}^\text{LO}
+\sigma_{L,T}^{qg, \text{sub.*}}
+\sigma_{L,T}^\text{dip} \, ,
\end{equation}
where $\sigma_{L,T}^{qg, \text{sub.*}}$ is an approximation of Eq.~(\ref{eq:NLO_qg_sub}) by using $X(z_2)=\xbj$ and taking the limit $\xbj/x_0 \to 0$ in the lower limit of the integral over $z_2$. This is the analogue of the ``CXY'' subtraction scheme in the case of single inclusive particle production, which is formally equivalent at this order of perturbation theory, but leads to problematic reults for high momentum scales.

\begin{figure*}[tbp]
\includegraphics[scale=\figscale]{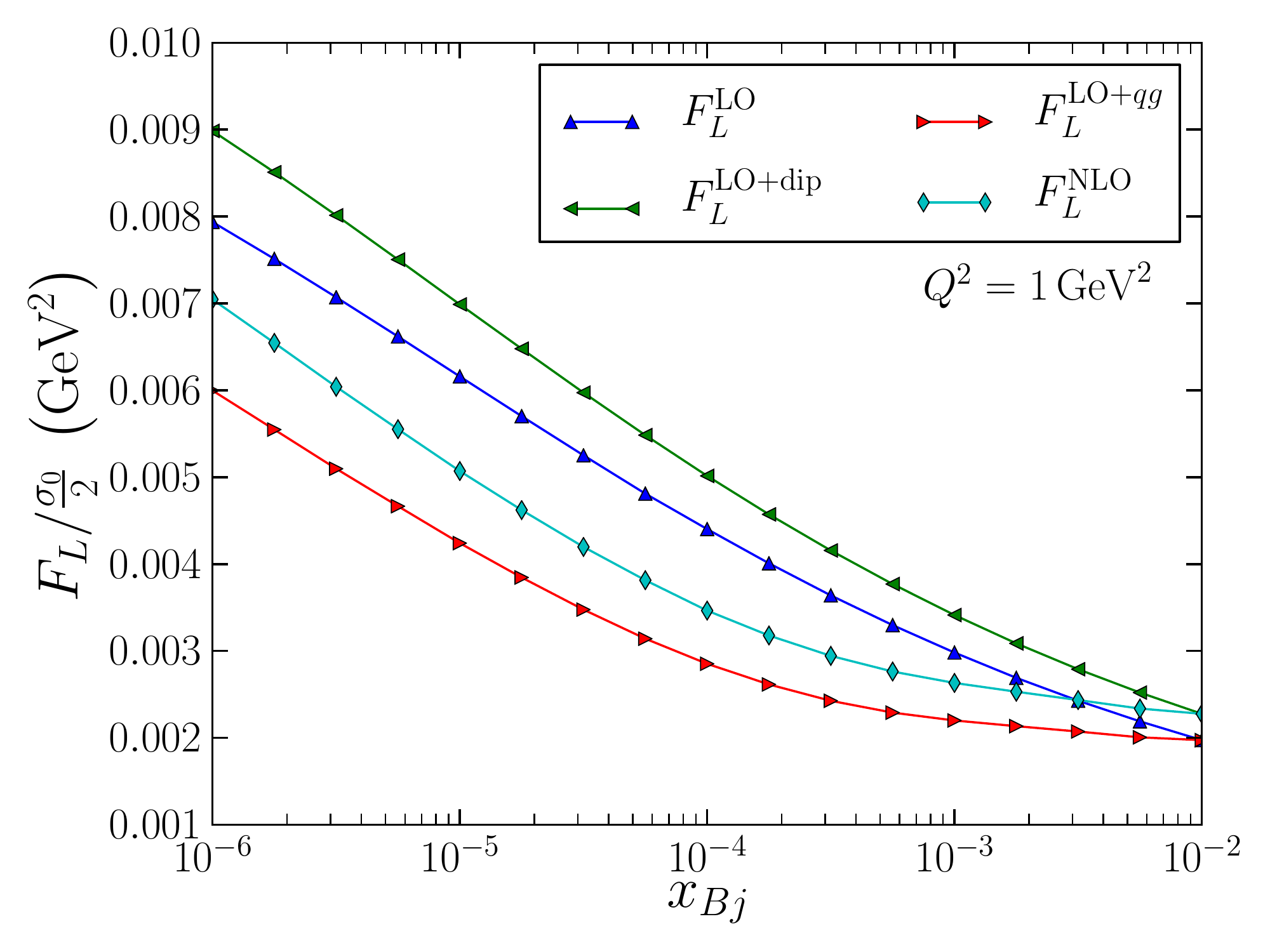}
\hspace{\figspace}
\includegraphics[scale=\figscale]{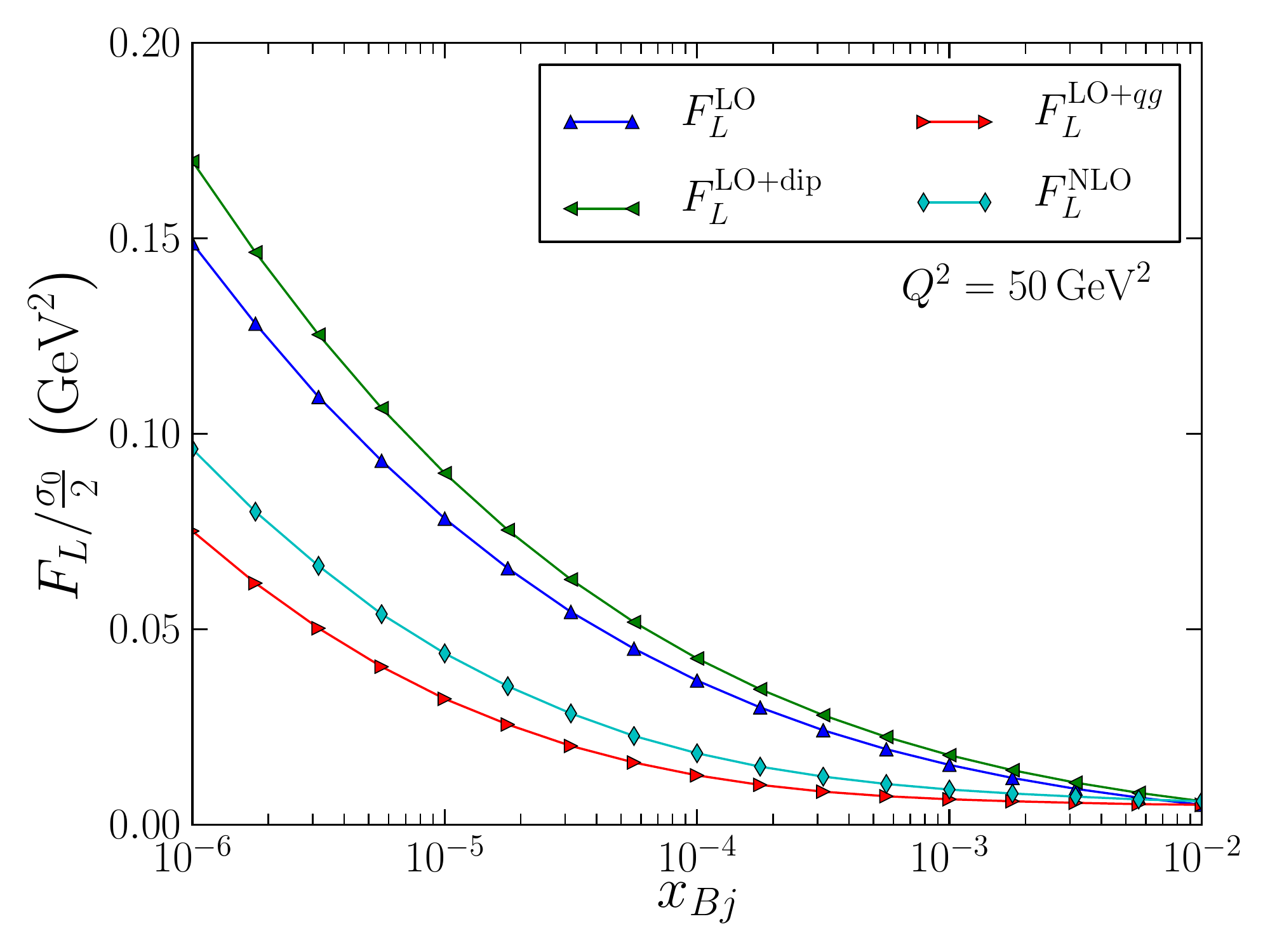}
\caption{LO and NLO contributions to $F_L$ as a function of $\xbj$ at $Q^2=1$ GeV$^2$ (left) and $Q^2=50$ GeV$^2$ (right) with $\as=0.2$. }
\label{fig:fc_x_L}
\end{figure*}

\begin{figure*}[tbp]
\includegraphics[scale=\figscale]{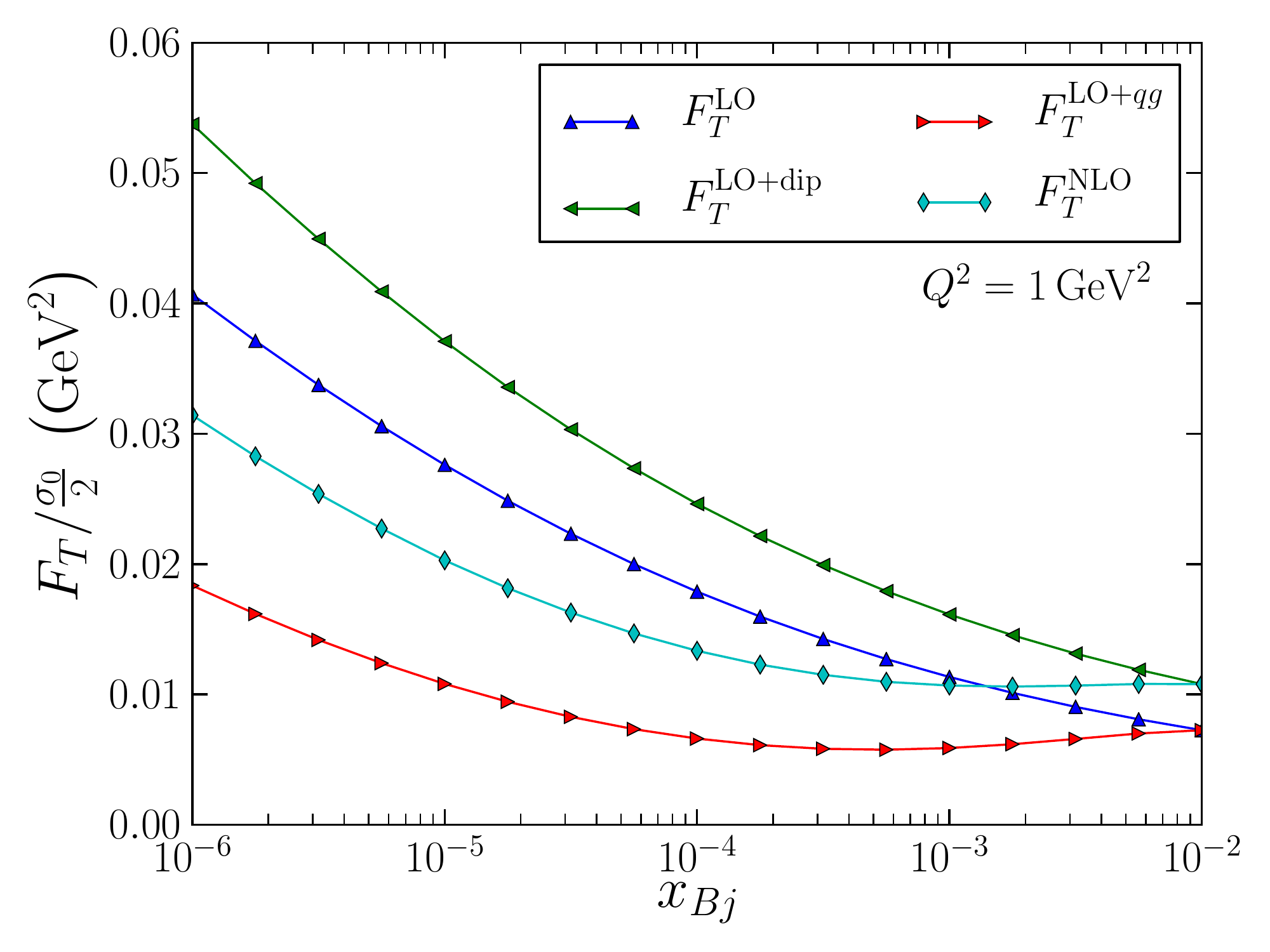}
\hspace{\figspace}
\includegraphics[scale=\figscale]{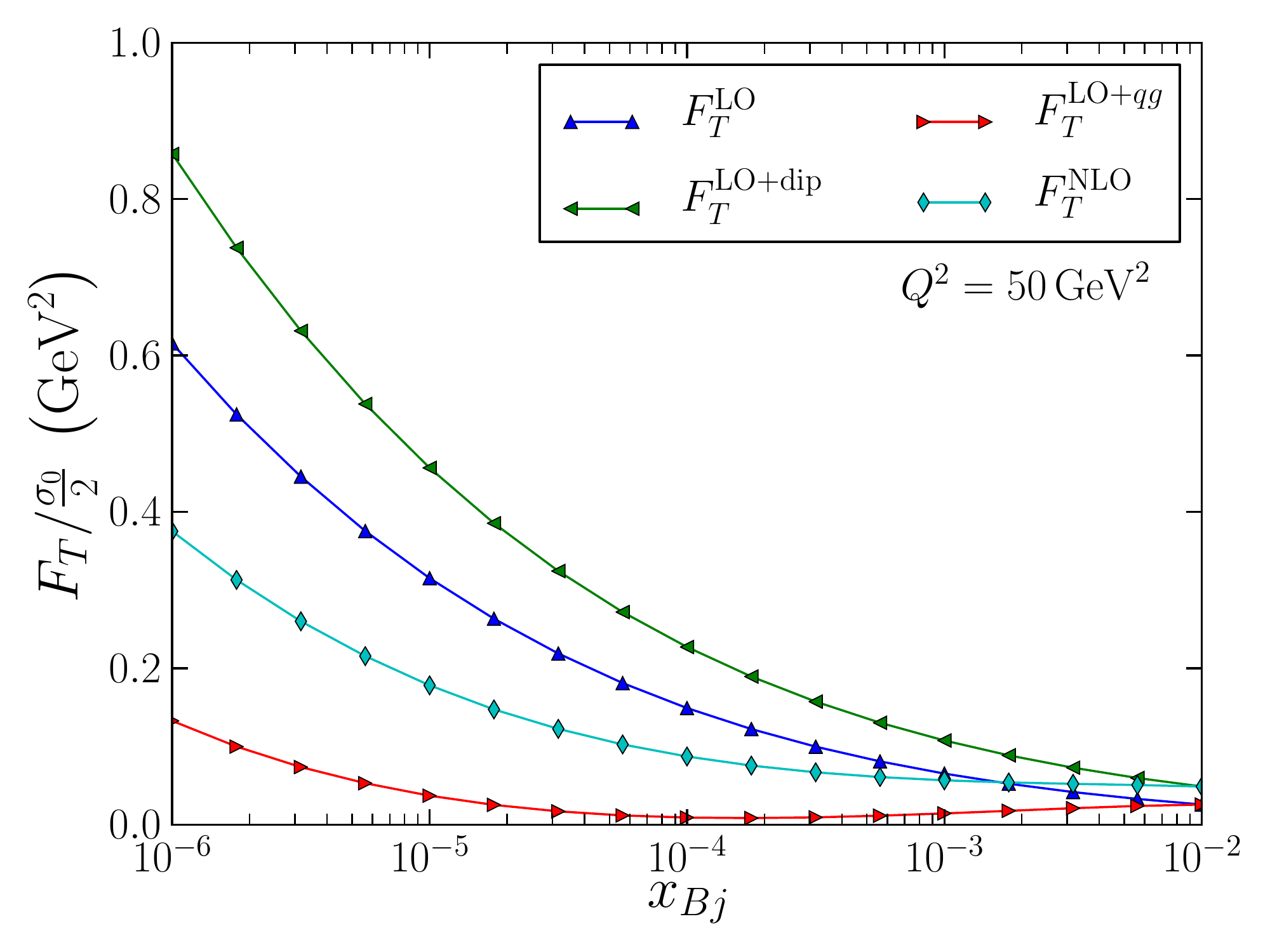}
\caption{LO and NLO contributions to $F_T$ as a function of $\xbj$ at $Q^2=1$ GeV$^2$ (left) and $Q^2=50$ GeV$^2$ (right) with $\as=0.2$.}
\label{fig:fc_x_T}
\end{figure*}

\begin{figure*}
\includegraphics[scale=\figscale]{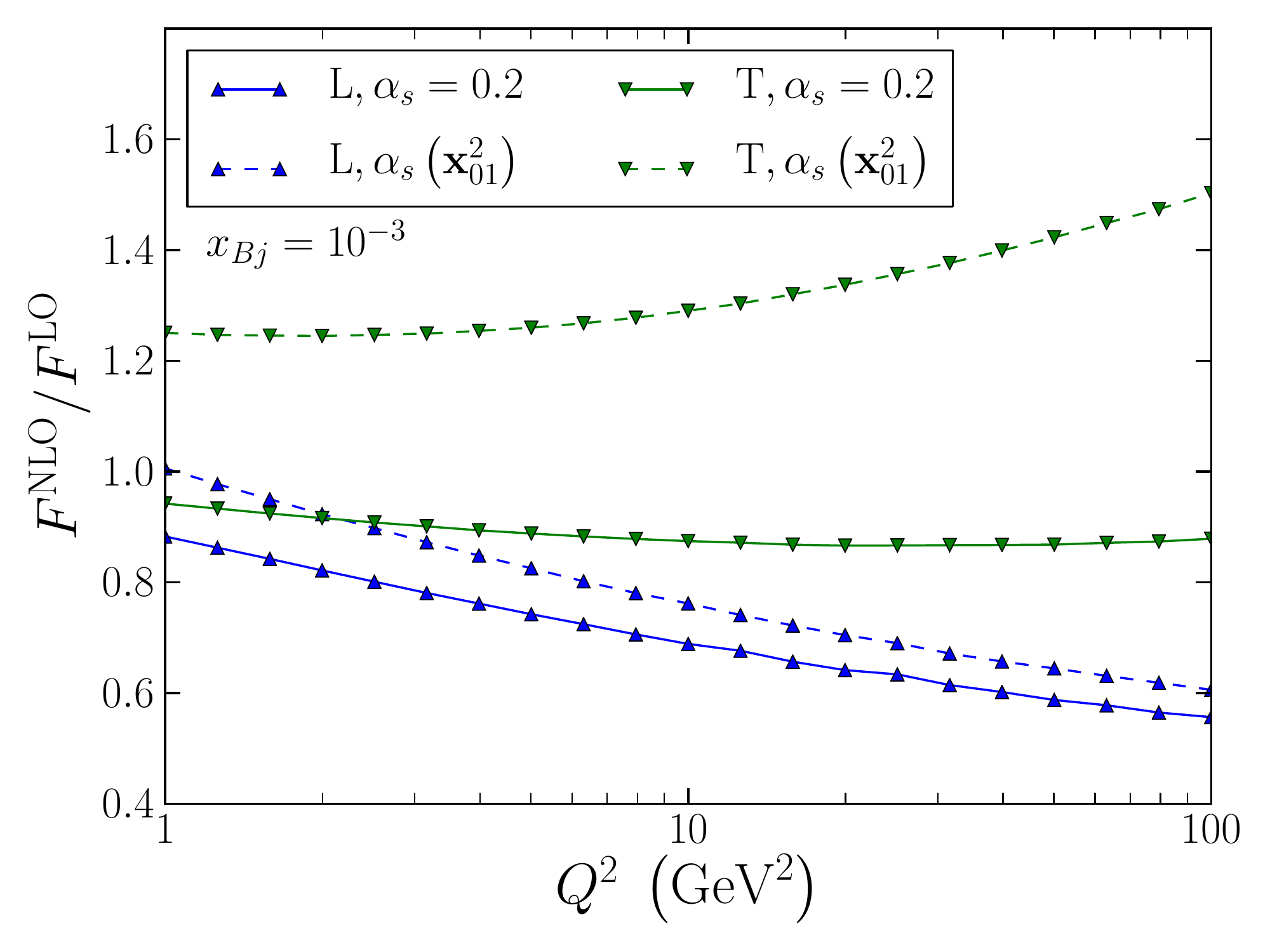}
\hspace{\figspace}
\includegraphics[scale=\figscale]{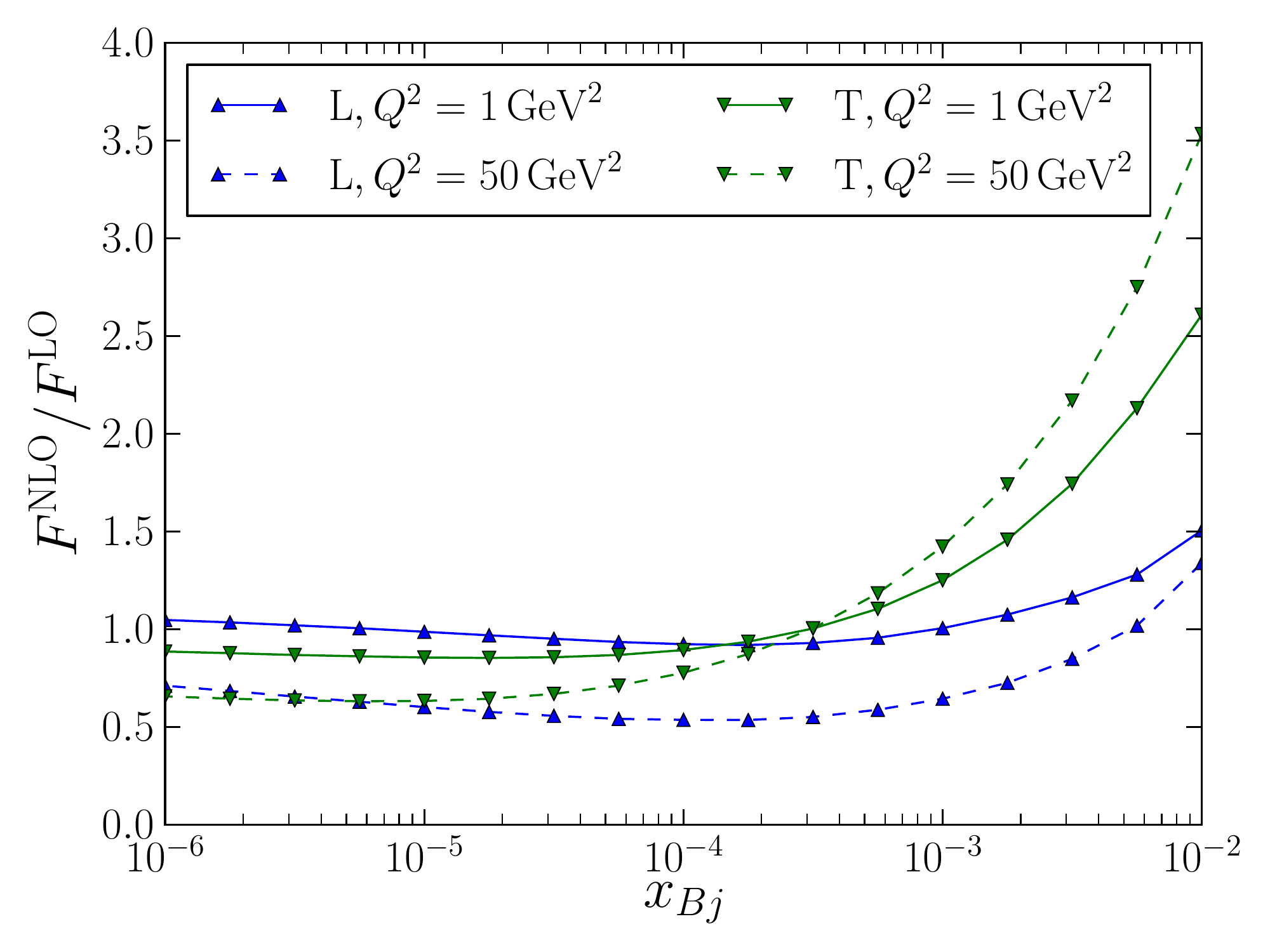}
\caption{Left: NLO/LO ratio for $F_L$ and $F_T$ as a function of $Q^2$ at $\xbj=10^{-3}$ with fixed (solid) and running (dashed) coupling. Right: NLO/LO ratio for $F_L$ and $F_T$ as a function of $\xbj$ at $Q^2=1$ GeV$^2$ (solid) and $Q^2=50$ GeV$^2$ (dashed) with running coupling.}
\label{fig:rc}
\end{figure*}

\section{Numerical results}
\label{sec:results}

Since we do not consider a possible impact parameter dependence of the dipole correlators, one of the coordinate integrals in the expressions shown in the previous section is trivial and leads to a factor corresponding to the target transverse area, denoted as $\sigma_0/2$. This quantity is usually determined by a fit to data, such as in~\cite{Albacete:2010sy,Lappi:2013zma}. Performing such a fit goes well beyond the scope of the present work, therefore for simplicity we leave out this overall normalization factor and present results for $F_{L,T}/\frac{\sigma_0}{2}$, where the structure functions $F_{L,T}$ are defined as
\begin{equation}
F_{L,T}(\xbj,Q^2)=\frac{Q^2}{4\pi^2\alpha_\text{em}}\sigma_{L,T}(\xbj,Q^2).
\end{equation}

We first focus on the fixed coupling case, using $\as=0.2$ both when evaluating the NLO cross section and when solving the leading order Balitsky-Kovchegov equation. Note that for the factorization scheme to be consistent both the cross section calculation and the BK equation need to have the same coupling constant. For the BK equation we use an MV initial condition~\cite{McLerran:1993ni}
\begin{equation}
	S(\rt,x_0)=\exp\left[-\frac{\rt^2 \qso^2}{4}\ln{\left(\frac{1}{|\rt| \lqcd}+e\right)}\right],
\end{equation}
where we take $\qso^2=0.2$ GeV$^2$ and $\lqcd=0.241$ GeV.

In Fig.~\ref{fig:fc_Q} we show the importance of the NLO corrections $\sigma^\text{dip}$ and $\sigma^{qg}$ to $F_L$ and $F_T$ as a function of $Q^2$ at $\xbj=10^{-3}$. In both the longitudinal and transverse cases the sign of these corrections is the same: the dipole contribution is positive, which can be understood from Eq.~(\ref{eq:NLO_dip}), while the $qg$ contribution is negative. Because the second correction is larger in magnitude than the first one, the total NLO cross section is smaller than the LO one.

In Fig.~\ref{fig:fc_Q_sub} we show how these results change if we use the approximate $\xbj$-subtraction in \eq\nr{eq:NLO_xbjsub} for the $qg$ term. This term is still negative and has a larger magnitude, especially at large $Q$, which makes the whole NLO cross section negative for $Q^2 \gtrsim 10$ GeV$^2$, both in the longitudinal and transverse cases. Therefore, approximating \eq\nr{eq:NLO_sub} by \eq\nr{eq:NLO_xbjsub}, while in principle justified in a weak coupling sense, has in fact a large effect in this region and can lead to unphysical results. A similar behavior was observed in single inclusive particle production at large transverse momenta~\citep{Ducloue:2017mpb}. This shows that to get meaningful results one should really use the factorization procedure in \eq\nr{eq:NLO_unsub} or equivalently \eq\nr{eq:NLO_sub}, which we will do for the rest of this paper.

We  also show in Figs.~\ref{fig:fc_x_L} and~\ref{fig:fc_x_T} the $\xbj$-dependence of the different NLO contributions to $F_L$ and $F_T$ for fixed $Q^2=1$ and 50 GeV$^2$. These plots show a change of behavior: at small $\xbj$ the NLO cross section is smaller than the LO one, while it becomes larger when $\xbj$ approaches $x_0$. The reason is the following: as explained previously, the dipole NLO correction is always positive. In addition, as can be seen from Eq.~(\ref{eq:NLO_qg_unsub}), the $qg$ part is 0 at $\xbj=x_0$ since the $z_2$-integration range vanishes. Therefore the NLO cross section is the sum of the leading order one and a positive correction, i.e. always larger than the leading order one. This is related to the reason why, as explained in the previous section, we would prefer to use an expression of the dipole part which has an explicit integration over $z_2$. This would allow one to use, also in the ``dipole'' term,  Wilson line operators at a  rapidity scale which depends on the gluon momentum fraction, i.e. the invariant mass of the $q\bar{q}g$-state, in a way that is more consistent with the $qg$ part. The expressions we currently use restrict the kinematics to the regime of validity of the dipole picture $X<x_0$ for the  $qg$-part, but not for the dipole part. This leads to a sign change of the total NLO contribution as a function of $\xbj$ near $x_0$.

While the running of the strong coupling $\as$ is in principle a subleading effect in a leading order calculation, this effect has to be taken into account at next-to-leading order. To evaluate its importance here, we use the simple parent dipole prescription  in which the coupling is given by
\begin{equation}
\as(\xt_{01}^2)=\frac{4\pi}{\beta_0 \ln\left(\frac{4 C^2}{\xt_{01}^2 \lqcd^2}\right)} \, ,
\end{equation}
with $\beta_0=(11\nc-2\nf)/3$. The scaling parameter $C^2$ is taken to be $C^2=e^{-2\gamma_e}$, as suggested in Refs.~\cite{Kovchegov:2006vj,Lappi:2012vw}, and the coupling is frozen at the value 0.7 at large dipole sizes. When fitting the initial condition of the BK equation to data at leading order (see e.g.~\cite{Albacete:2010sy,Lappi:2013zma}), one usually uses instead the Balitsky prescription~\cite{Balitsky:2006wa} for the running coupling and additionally takes $C^2$ as a fit parameter in order to obtain a slow enough evolution. However, in principle the choice of the running coupling prescription is a higher order effect, and thus the parent dipole prescription is equally well justified in a weak coupling sense. Also on the phenomenological level it has been shown~\cite{Lappi:2015fma,Iancu:2015vea,Iancu:2015joa,Lappi:2016fmu} that the NLO corrections to the BK kernel slow down the evolution, and thus it is not a priori obvious which prescription will yield a good description of experimental data at the NLO level.

As stated before, our purpose here is not to achieve a fit to DIS data, but to quantify the effect of the NLO corrections to the impact factor compared to previous LO calculations. Therefore we show, in the left panel of Fig.~\ref{fig:rc}, the NLO/LO ratio for $F_L$ and $F_T$ as a function of $Q^2$ at $\xbj=10^{-3}$ with fixed and running coupling. In the right panel we show the same ratio as a function of $\xbj$ at $Q^2=1$ and 50 GeV$^2$ with running coupling. We see that for fixed coupling, the net effect of the NLO corrections is to decrease the cross section. However, especially for a running coupling, this feature is reversed close to the initial rapidity scale $\xbj \approx x_0$. As discussed above, this is related to the fact that the negative NLO corrections related to BK evolution vanish in this limit while the positive ones in the ``dipole'' term do not, indicating a strong dependence on the details of the factorization scheme. While this is a transient effect that does not alter the asymptotic high energy behavior, treating it carefully will be important for an attempt to describe experimental data.

\section{Outlook}

In conclusion, we have in this paper evaluated, for the first time, the total DIS cross section in the dipole picture with an impact factor derived at NLO accuracy. We developed a factorization procedure to resum the leading high energy logarithms into a BK renormalization group evolution of the target, in line with recent developments for single inclusive cross sections. We showed that this procedure leads to physical, well-behaved expressions for the cross sections with, however, large transient effects in the region close to the limit of validity of the eikonal approximation. With the caveat of understanding these transient effects, there is a good perspective for a comparison with experimental data. In order to achieve this at consistent NLO accuracy, the impact factors studied here must be combined with a solution of the NLO BK equation~\cite{Lappi:2016fmu} or at least a collinearly resummed version of the LO equation~\cite{Iancu:2015vea,Iancu:2015joa}. A major missing theoretical ingredient that is needed for a more detailed comparison with data is to work out the corresponding impact factor for massive quarks. This should in principle be a straightforward, if laborious, extension of the existing calculation for massless quarks.

\section*{Acknowledgments} 
We thank G. Beuf for sharing and discussing the results of \cite{Beuf:2017bpd} before publication, R. Paatelainen for discussions and H. Mäntysaari for sharing his BK evolution code. This work has been supported by the Academy of Finland, projects 273464 and 303756 and by the European Research Council, grant ERC-2015-CoG-681707.

\bibliographystyle{JHEP-2modlong}
\bibliography{spires}

\end{document}